\documentstyle[12pt]{article}

\catcode`\@=11
\@addtoreset{equation}{section}

\catcode`\@=12

\begin{document}

\title{Is Neutrino a Superluminal Particle?}
\author{Guang-jiong NI\thanks{%
Department of Physics, Fudan University, Shanghai, 200433, China}
$\quad$ and Tsao CHANG\thanks{%
Center for Space Plasma and Aeronomy Research, University of Alabama in
Huntsville, Huntsville, AL 35899, USA} }
\date{}
\maketitle

\begin{abstract}
Based on the experimental discovery that the mass-square of neutrino is
negative, a quantum theory for superluminal neutrino is proposed. Two Weyl
equations coupled together via a mass term respecting the maximum parity
violation lead to a new equation which describes the superluminal motion of
neutrino with permanent helicity. Various strange features of subluminal and
superluminal particles can be ascribed to the relative variation of two
contradictory fields superposing coherently inside the particle with the
change of its speed u in the whole range ($0<u<\infty$). Being compatible
with the theory of special relativity, this theory may have various
applications. \vskip 5mm

\end{abstract}

\vskip 1cm \baselineskip=7mm

Recent experimental data in the measurement of tritium beta decay reveal an
amazing result that the mass-square of electron neutrino is negative [1]: 
$$
m^2(\nu _e)=-2.5\pm 3.3eV^2.\eqno{(1)} 
$$
The pion decay experiment also shows a similar puzzle that [1]: 
$$
m^2(\nu _\mu )=-0.016\pm 0.023MeV^2.\eqno{(2)} 
$$
The experimental data (1) and (2),though far from accurate, strongly hint
that a neutrino might be some particle moving faster than light. Following
the existing literature [2,3,4,5,6], we shall call it superluminal particle
or tachyon, which obeys the kinematic relation: 
$$
E^2=c^2p^2-m_s^2c^4\eqno{(3)} 
$$
with its energy, momentum and ``proper mass'' denoted by $E,\;p$ and $m_s$
respectively, For instance $m_s(\nu _e)=1.6eV$. In a short letter [7] a new
Dirac-type equation is established for explaining the negative mass-square
of neutrino. As a continuation and development of [7], here we will
elaborate the quantum theory in some detail and explore further the
intrinsic essence responsible for the strange behavior of tachyon.

Since the discovery of parity violation in 1956 [8, 9], the theory for
neutrino is based on Weyl equation 
$$
i\hbar \frac \partial {\partial t}\xi =ic\hbar \vec{\sigma}\cdot \nabla \xi ,%
\eqno{(4)}
$$
where $\xi (t,x)$ is a two-component spinor function. Eq. (4) describes a
positive energy $(E>0)$ solution for left-handed neutrino (with helicity $H=<%
\vec{\sigma}\cdot \widehat{\vec{p}}>=-1,\;\widehat{\vec{p}}=\vec{p}/|\vec{p}|
$) and a negative energy ($E<0$) solution for right-handed antineutrino
(with helicity $H=1$) in accordance with that verified by experiments. The
alternative possibility that 
$$
i\hbar \frac \partial {\partial t}\eta =-ic\hbar \vec{\sigma}\cdot \nabla
\eta ,\eqno{(5)}
$$
was thus abandoned. As now experiments show that $m_s\neq 0$ , we assume a
new equation for neutrino being composed of both $\xi $ and $\eta $ coupling
via nonzero $m_s$: 
$$
\begin{array}{l}
i\hbar \frac \partial {\partial t}\xi =ic\hbar \vec{\sigma}\cdot \nabla \xi
-m_sc^2\eta ,\nonumber \\ 
i\hbar \frac \partial {\partial t}\eta =-ic\hbar \vec{\sigma}\cdot \nabla
\eta +m_sc^2\xi 
\end{array}
\eqno{(6)}
$$

Consider a plane wave solution $\xi \sim \eta \sim \exp [i(px-Et)/\hbar ]$
along $x$ axis for a particle with helicity $H=-1$, we find 
$$
\eta =\frac{m_sc^2}{cp+E}\xi \eqno{(7)} 
$$
and Eq.(3) as expected. Based on Eq.(3) with quantum relations $E=\hbar
\omega $ and $p=\hbar k$, the velocity of particle, $u$, should be
identified with the group velocity $u_g=\frac{d\omega }{dk}$ of wave versus
the phase velocity $u_p=\frac \omega k$. Defining the changeable (total)
mass $\tilde{m}$ by $p=\tilde{m}u_g=\tilde{m}u$, one can easily prove that: 
$$
u_pu_g=c^2,\eqno{(8)} 
$$
$$
p=\tilde{m}u=\frac{m_su}{\sqrt{\frac{u^2}{c^2}-1}},\quad E=\tilde{m}c^2=%
\frac{m_sc^2}{\sqrt{\frac{u^2}{c^2}-1}}.\eqno{(9)} 
$$

For understanding why a negative solution ($E<0$) describes an antiparticle,
we introduce two linear combination functions of $\xi$ and $\eta$ : 
$$
\varphi=\frac{1}{\sqrt{2}}(\xi +\eta ),\quad \chi =\frac{1}{\sqrt{2}}(\xi
-\eta ) \eqno{(10)} 
$$
$$
\begin{array}{l}
i\hbar \frac{\partial}{\partial t}\varphi = ic\hbar\vec{\sigma}%
\cdot\nabla\chi +m_sc^2\chi,\nonumber \\ 
i\hbar \frac{\partial}{\partial t}\chi = ic\hbar\vec{\sigma}%
\cdot\nabla\varphi-m_sc^2\varphi
\end{array}
\eqno{(11)} 
$$
Eq. (11) is invariant under the space-time inversion ($\vec{x}%
\longrightarrow -\vec{x}$, $t\longrightarrow -t$) and transformation: 
$$
\varphi (-\vec{x},-t)\longrightarrow \chi (\vec{x},t),\quad \chi (-\vec{x}%
,-t)\longrightarrow \varphi (\vec{x},t)\eqno{(12)} 
$$
(see Refs. [10,11]). For a concrete solution of particle: 
$$
\varphi\sim\chi\sim\exp[i(px-Et)/\hbar],\quad (E>0)\eqno{(13)} 
$$
$$
\chi=\frac{cp-m_sc^2}{E}\varphi, (|\varphi/\chi |>1)\eqno{(14)} 
$$
the transformation (12) leads to the wavefunction (WF) of its antiparticle: 
$$
\begin{array}{l}
\varphi (-\vec{x},-t)\longrightarrow \chi_c (\vec{x},t)\sim\exp[%
-i(px-Et)/\hbar], \\ 
\chi (-\vec{x},-t)\longrightarrow \varphi_c (\vec{x},t)\sim\exp[%
-i(px-Et)/\hbar],
\end{array}
\quad (|\chi_c/\varphi_c |>1))\eqno{(15)} 
$$
Eq.(15) implies a negative energy solution if we use the familiar operators
of momentum and energy for particle: 
$$
\hat{p}=-i\hbar\frac{\partial}{\partial x},\quad \hat{E}=i\hbar\frac{\partial%
}{\partial t}\eqno{(16)} 
$$
But for antiparticle, we should use the counterparts of (16) as [11] 
$$
\hat{p_c}=i\hbar\frac{\partial}{\partial x},\quad \hat{E_c}=-i\hbar\frac{%
\partial}{\partial t}\eqno{(16)} 
$$
(subscript $c$ denotes the antiparticle). Hence Eq. (15) describes an
antiparticle with momentum p and energy $E(>0)$ precisely as that in Eq.
(13) for a particle. Interesting enough, a modification of Eq. (11) as 
$$
\begin{array}{l}
i\hbar \frac{\partial}{\partial t}\varphi_D = ic\hbar\vec{\sigma}%
\cdot\nabla\chi_D+m_0c^2\varphi_D,\nonumber \\ 
i\hbar \frac{\partial}{\partial t}\chi_D = ic\hbar\vec{\sigma}%
\cdot\nabla\varphi_D -m_0c^2\chi_D
\end{array}
\eqno{(18)} 
$$
which still obeys the symmetry (12) is just the Dirac equation: 
$$
i\hbar \frac{\partial}{\partial t}\psi_D = ic\hbar\vec{\alpha}%
\cdot\nabla\psi_D+\beta m_0c^2\psi_D\eqno{(19)} 
$$
$$
\psi_D=\left( 
\begin{array}{c}
\varphi_D \\ 
\chi_D
\end{array}
\right),\quad \alpha_i= \left( 
\begin{array}{cc}
0 & \sigma_i \\ 
\sigma_i & 0
\end{array}
\right),\quad \beta= \left( 
\begin{array}{cc}
I & 0 \\ 
0 & -I
\end{array}
\right)\eqno{(20)} 
$$
for describing the electron with rest mass $m_0$ and speed $u<c$.

Why Eq.(11) is so radically different from the Eq.(18)? Let's first derive
the continuity equation for Eq. (11) (or Eq.(6)): 
$$
\frac{\partial \rho }{\partial t}+\nabla \cdot \vec{j}=0,\eqno{(21)} 
$$
$$
\rho =\varphi ^{+}\chi +\chi ^{+}\varphi =\xi ^{+}\xi -\eta ^{+}\eta ,%
\eqno{(22)} 
$$
$$
\vec{j}=-c(\varphi ^{+}\vec{\sigma}\varphi +\chi ^{+}\sigma \chi )=-c(\xi
^{+}\vec{\sigma}\xi +\eta ^{+}\vec{\sigma}\eta ),\eqno{(23)} 
$$
versus that for Dirac Eq.(18) or its equivalent form: 
$$
\xi _D=\frac 1{\sqrt{2}}(\varphi _D+\chi _D),\quad \eta _D=\frac 1{\sqrt{2}%
}(\varphi _D-\chi _D)\eqno{(24)} 
$$
$$
\begin{array}{l}
i\hbar \frac \partial {\partial t}\xi _D=ic\hbar \vec{\sigma}\cdot \nabla
\xi _D+m_0c^2\eta _D,\nonumber \\ 
i\hbar \frac \partial {\partial t}\eta _D=-ic\hbar \vec{\sigma}\cdot \nabla
\eta _D+m_0c^2\xi _D
\end{array}
\eqno{(25)} 
$$
$$
\rho _D=\varphi _D^{+}\varphi _D+\chi _D^{+}\chi _D=\xi _D^{+}\xi _D+\eta
_D^{+}\eta _D,\eqno{(26)} 
$$
$$
\vec{j}_D=-c(\varphi _D^{+}\vec{\sigma}\chi _D+\chi _D^{+}\vec{\sigma}%
\varphi _D)=-c(\xi _D^{+}\vec{\sigma}\xi -\eta _D^{+}\vec{\sigma}\eta _D).%
\eqno{(27)} 
$$

Now the normalization condition $\int\rho d\vec{x}=1$ for Eq.(6) corresponds
to the conservation of helicity in the motion: $H= -1$ for particle whereas $%
H= 1$ for antiparticle. The neutrino (antineutrino) is permanently
longitudinal polarized as $\nu_L(\bar{\nu}_R)$ and its invariant feature can
be maintained in any inertial frame because of its velocity $u>c$.

Next, we can find out a radical difference between (6) and (25). Under the
space-inversion ($\vec{x}\longrightarrow -\vec{x}$) and related
transformation: 
$$
\xi_D (-\vec{x},t)\longrightarrow \eta_D (\vec{x},t),\quad \eta_D (-\vec{x}%
,t)\longrightarrow \xi_D (\vec{x},t)\eqno{(28)} 
$$
Dirac Eq.(25) is invariant whereas Eq. (6) fails to do so because of the
opposite sign in mass term. It is just a clearcut reflection of the fact
that neutrino yields the maximum violation of parity. The new observation is
that the parity violation is triggered by its nonzero (proper) mass which in
turn implies that neutrino must be a superluminal particle with permanent
helicity: while $\nu_L$ and $\bar{\nu}_R$ are allowed, $\nu_R$ and $\bar{\nu}%
_L$ must be forbidden strictly.

We are now in a position to realize the marvelous kinematical feature of
superluminal particle together with that of subluminal one. We define the
``ratio'' R of ``hidden amplitude of antiparticle state'' to that of
``particle state'' in a particle (as what had been done for Dirac particle
in Ref. [12] with R there being redefined as R$^2$ here): 
$$
R\equiv \sqrt{\frac{\chi ^{+}\chi }{\varphi ^{+}\varphi }}=\left[ \frac{%
\frac uc-\sqrt{\frac{u^2}{c^2}-1}}{\frac uc+\sqrt{\frac{u^2}{c^2}-1}}\right]
^{\frac 12},\quad (u>c)\eqno{(29)} 
$$
$$
R=\left[ \frac{1-\sqrt{1-\frac{u^2}{c^2}}}{1+\sqrt{1-\frac{u^2}{c^2}}}%
\right] ^{\frac 12},\quad (u<c).\eqno{(30)} 
$$

Similarly, we define a ``Weyl parameter'' W as the ratio of ``hidden
amplitude of right-handed helicity state'' to that of ``left-handed helicity
state'' in a particle with helicity $H=-1$: 
$$
W=\sqrt{\frac{\eta ^{+}\eta }{\xi ^{+}\xi }}=\sqrt{\frac{u-c}{u+c}},\quad
(u>c)\eqno{(31)} 
$$
$$
W=\sqrt{\frac{c-u}{c+u}},\quad (u<c).\eqno{(32)} 
$$

Being functions of ($u/c$), the values of $R$ and $W$ are symmetric with
respect to $u/c=1$ in logarithmic scale. So we define the ``rapidity'' $y$ of
particle with $u$ in whole range for both $u<c$ and $u>
c$:
$$
y=\ln \sqrt{\left| \frac{c+u}{c-u}\right| }\eqno{(33)} 
$$
and find that both R and W can be expressed in a unified manner:
$$
R=tanh\left( \frac y2\right) ,W=\exp \left( -y\right) \eqno{(34)} 
$$
which in turn are anticorrelated each other also in a unified way:

$$
R=\frac{1-W}{1+W},W=\frac{1-R}{1+R}\eqno{(35)} 
$$

The mysteries of SR are now unveiled. The answer is ascribed to the
monotonical increase of ``hidden field of antiparticle state'' and its phase
evolution being opposite to that of particle state essentially as shown by
Eq.(15) versus (13). Though due to the condition $|\varphi /\chi |>1$, the
``hidden antiparticle field'' $\chi $ is in subordinate position and is
subjected to follow the ``particle field'' $\varphi $ as shown in Eq.(13),
it does impose an opposite tendency and enhance the inertial mass ($\tilde{m}
$) of particle. In some sense, the time reading of clock accompanying $%
\varphi $ is clockwise whereas that of $\chi $ is anticlockwise essentially.
Though the time reading of a moving clock remains clockwise, it runs slower
and slower with the enhancement of $\chi $ field.

Once the speed limit for a subluminal particle, the speed of light $c$, is
broken through, a superluminal particle emerges with even more mysterious
behavior as shown in Eqs. (3) and (9). Why they are so radically different
from that of subluminal particle? The reason can be found from the
difference between Eqs. (22) and (26) together with behavior of Weyl
parameter. The normalization condition for Dirac particle $\int\rho_Dd\vec{x}
$ imposes stringent constraint on $\xi_D$ and $\eta_D$ such that $%
\displaystyle{\lim_{u\longrightarrow 0}}\int\xi^+_D\xi_Dd\vec{x}=%
\displaystyle{\lim_{u\longrightarrow 0}}\int\eta^+_D\eta_Dd\vec{x}=\frac 12$%
. By contrast, the normalization condition for superluminal particle $%
\int\rho d\vec{x}=1$ imposes no constraint on $\xi$ and $\eta$ separately.
The limiting behavior of $\displaystyle{\lim_{u\longrightarrow \infty}}W=1$
implies that both of them increase infinitely and approach to equal strength
when $u\longrightarrow \infty$. The fantastic consequence is their
cancellation effect on the hidden antiparticle field $\chi=\frac{1}{\sqrt{2}}%
(\xi -\eta)$ to render $R\longrightarrow 0$ and hence the tachyon energy $E=%
\tilde{m}c^2\longrightarrow 0$ while momentum $p=\tilde{m}u\longrightarrow
m_sc$.

\vskip 4mm

{\bf {\large Summary and discussion:}}

(a) Based on QM, a theory for superluminal particle is established. Being
compatible with the theory of SR, it's actually a complement to SR. Now we
realize that a particle can have speed u varying in the whole range ($%
0,\infty$) with a universal constant c (the speed of light ) as a
singularity dividing superluminal particle (tachyon) from subluminal one.
Most likely, the neutrino is just a tachyon with spin $1/2$.

(b) The crucial point is the cognition that ``a particle is always not pure''
[10-12] and there is no exception to neutrino. Now we realize that once if
neutrino has some mass, no matter how tiny it is, two Weyl equations, (4)
and (5), should be coupled together via some mass term while still
respecting the maximum parity violation. Then Eq. (6) emerges almost as a
unique possibility and an inevitable conclusion turns out to be that
neutrino must be a superluminal particle with permanent helicity.

(c) Rewriting Eq. (11) in the form of four-component spinor equation, we
find a Dirac-type equation [7]: 
$$
i\hbar \frac \partial {\partial t}\psi _s=ic\hbar \vec{\alpha}\cdot \nabla
\psi _s+\beta _sm_sc^2\psi _s,\eqno{(36)}
$$
$$
\psi _s=\left( 
\begin{array}{c}
\varphi  \\ 
\chi 
\end{array}
\right) ,\quad \alpha _i=\left( 
\begin{array}{cc}
0 & \sigma _i \\ 
\sigma _i & 0
\end{array}
\right) ,\quad \beta _s=\left( 
\begin{array}{cc}
0 & I \\ 
-I & 0
\end{array}
\right) .\eqno{(37)}
$$

However, in comparison with Dirac Eq.(19), $\beta_s$ is not a hermitian
matrix. Now we realize that the violation of hermitian property is stemming
from the violation of parity. Though a nonhermitian Hamiltonian is not
allowed for a subluminal particle because it would lead to instability
of solutions, it does work for a superluminal particle. Of four
solutions for a same momentum of neutrino, two of them are eliminated
due to the parity violation, corresponding to $\nu_R$ and
$\bar{\nu}_L$ being forbidden strictly, and other two are stabilized,
corresponding to physical realization of $\nu_L$ and
$\bar{\nu}_R$. More importantly, Eq. (36) still preserves the
invariance of basic symmetry (12).

(d) The parameter R defined in Eq.(29) could be understood as a measure of
``impurity'' of a particle being a superposition state of two hidden
contradictory fields $\varphi$ and $\chi (|\varphi /\chi|>1)$. Though
superficially, a free electron (neutrino) is always a particle with lepton
quantum number $L=1$, it does change intrinsically with its velocity. The
larger $R$ is, the larger mass(energy) and more instability it will have.

Similarly, the Weyl parameter $W$ is the measure of ``intrinsic instability
of helicity'' of a particle with superficial helicity $H=-1$, ($|\xi/\eta|>1$%
). While an electron can trun its helicity to $H=1$ when $|\eta/\xi|>1$, a
neutrino's helicity is linked to lepton number $L$ ($L=1,\; H=-1$ whereas $%
L=-1,\; H=1$) definitely. This difference is stemming from Eq.(26) versus
(22). The common anticorrelation between $R$ and $W$ for both subluminal and
superluminal particles implies that a high energy particle being more
``impure'' ($R\longrightarrow 1$) will be more stable in helicity ($%
W\longrightarrow 0$). On the contrary, a particle being unstable in helicity
($W\longrightarrow 1$) corresponds to a ``relatively pure'' particle ($%
R\longrightarrow 0$) with low energy. A prominent difference between a Dirac
particle and neutrino is that for the former $\displaystyle{%
\lim_{u\longrightarrow 0}}E=m_0c^2$ whereas for the latter $\displaystyle{%
\lim_{u\longrightarrow \infty}}E=0$.

(e) When we talk about $\chi $ being the ``hidden antiparticle amplitude''
inside a particle and $\eta $ being the ``hidden right-handed helicity
amplitude'' inside a left-handed neutrino, we have to be cautious. As an
example, for a high-energy electron with $R=1/3$, can we say that ``it is
composed of $75\%$ (or 90\%)electron ingredient and $25\%$ (or
10\%)antielectron (positron) ingredient''? No, we can't. Because the $\chi $
field inside an electron is a probability amplitude and is in a subordinate
position, no hidden opposite ``charge'' can be observed in a high-energy
electron. The hidden $\chi $ field can only exhibit its implicit presence
via the strange SR effects. Similarly, for a neutrino with $W=1/3$, we can
not say that ``it is composed of $75\%$ (or 90\%) left-handed rotating state
and $25\%$ (or 10\%)right-handed rotating state'' because the neutrion is in 
$100\%$ left-handed helicity state explicitly while both $\xi $ and $\eta $
fields enhance drastically and cancel each other considerably inside. Only
in an antineutrino with $|\eta _c/\xi _c|>1$, can $\eta _c$ display itself
as a right-handed rotating state, so does $\xi _c$ follow accordingly.

Therefore, we should not interprete the ``hidden probability amplitude ''
too materialized in ordinary language. All fantastic behaviours of particle
are due to the linear superposition and interference effect of fields between $%
\varphi $ and $\chi $ (or $\xi $ and $\eta $ ), not due to their intensity ($%
\varphi ^{+}\varphi $ etc.). The existence of superluminal particle and its
marvelous feature are new manifestations of subtlety of QM .In some sense,a
particle is also a ``Schr\"{o}dinger's cat''in microscopic scale and could be
compared to the recent experimental verification of ``macroscopic
Schr\"{o}dinger's cat''[13] with its theoretical discussions [14,15].

(f) Consider a neutrino emitted in a supernova explosion. If being a
tachyon, it will arrive at our earth earlier than the visible light by a
time 
$$
\Delta T=\frac D{2c}\left( \frac{m_\nu c^2}E\right) ^2\eqno{(38)}
$$
with $D$ and $m_\nu (E)$ being the distance of supernova and proper
mass(energy) of neutrino repectively. On the other hand, the time span ($%
\Delta E$) in the arrival of neutrino will reflect its spreading in energy ($%
\Delta E$): 
$$
\Delta t\approx \frac Dc\left( \frac{\Delta u}c\right) \approx \frac
Dc\left( \frac{m_\nu c^2}E\right) ^2\frac{\Delta E}E\eqno{(39)}
$$
$$
\frac{\Delta t}{\Delta T}\approx 2\frac{\Delta E}E\eqno{(40)}
$$

In Feb.23, 1987, a supernova explosion (SN 1987A) was observed. Two
laboratories in Northern Hemisphere had detected $8$ and $12$ neutrino
events within a time interval $\Delta t\approx 6$ and $13s$ respectively
[16,17]. Since $D=1.6\times 10^5ly$, using the data in [16], we can estimate $%
\Delta u/c\sim 1.2\times 10^{-12}$ and $(m_\nu c^2)=44eV$ ( $E\sim 32
MeV$, $\Delta E\sim 20MeV)$, which might be regarded as some upper bound for
neutrino mass of all flavors. But more importantly, according to above
estimation, laboratories in Southern Hemisphere could observe the light
signal later than the neutrino events by a time lag $\Delta T\sim 5s$. It is
regrettable for having no original record available and the theory of
supernova explosion being too complicated for reliable conclusion could be
drawn from so limited data.

(g) If neutrino is really a tachyon with its motion equation shown as
Eq.(33) [i. e., Eq. (11) or (6)], then all the weak interaction processes in
which neutrino participates need to be restudied. We hope new clues might be
found for ``neutrino oscillation'' and the ``missing puzzle of solar neutrino''.
Especially, the mystery about dark matter in cosmos should be of the utmost
concern.


\begin{thebibliography}{99}
\bibitem{1}  ''Review of Particle Physics'', Euro. Phys. Journ. C 15 (2000)
350.

\bibitem{2}  O.M.P Bilaniuk et al, Am. J. Phys. 30 (1962) 718.

\bibitem{3}  G. Feinberg, Phys. Rev. 159 (1967) 1089.

\bibitem{4}  E.C.G. Sudarshan, in Proceeding of the VIII Nobel Symposium,
ed. by N. Swartholm (J. Wiley, New York, 1970),pp 335; J. Bandukwala and
D.Shay, Phys. Rev. D9 (1974) 889; D. Shay, Lett. Nuovo Cim. 19 (1977) 333;
E. Marx, Int. J. Theor. Phys. 3 (1970) 299.

\bibitem{5}  E. Recami et al, ''Tachyons, Monopoles and Related Topics'',
(North-Holland, 1978); A. Chodos et al. Phys. Lett. B 150 (1985) 431.

\bibitem{6}  T. Chang, J. Phys. A. 12 (1979) L203; in ``Proceedings of the
Sir A. Eddington Centenary Symposium, Vol. 3, Gravitatronal Radiation and
Relativity'' (1986) 431.J.Rembielinski,I.J.M.P.A12(1997)1677.

\bibitem{7}  T. Chang and G.-j. Ni, An explanation on negative mass-square
of neutrinos, Preprint, hep-ph/0009291.

\bibitem{8}  T.D. Lee and C.N. Yang, Phys. Rev. 104 (1956) 254.

\bibitem{9}  C.S. Wu et al, Phys. Rev. 105 (1957) 1413.

\bibitem{10}  G.-j. Ni and S.-q. Chen, J. Fudan University (Natural
Science), 35 (1996) 325.

\bibitem{11}  G-j. Ni, H. Guan, W.-m. Zhou and J. Yan, Chin. Phys. Lett. 17
(2000) 393.

\bibitem{12}  G.-j. Ni, W.-m. Zhou and J. Yan, in Proceeding of
International Workshop on Lorentz Group, CPT, and Neutrinos, Zacatecas,
Mexico, June 23-26, 1999 (Would Scientific. 2000).

\bibitem{13}  J.R. Friedman et al. Nature 406, (2000) 43.

\bibitem{14}  G. Blatter, Nature 406 (2000) 25.

\bibitem{15}  G.-j. Ni, What Schr\"{o}dinger's cat is telling, Preprint.

\bibitem{16}  R.M.Bionta et al. Phys. Rev. Lett.58 (1987)1494.

\bibitem{17}  K.Hirata et al. Phys. Rev. Lett. 58 (1987) 1490.
\end{thebibliography}
\end{document}